\documentclass{article}

\usepackage{microtype}
\usepackage{graphicx}
\usepackage{subfig}
\usepackage{booktabs} 
\usepackage{enumerate}
\usepackage{multicol}

\usepackage{hyperref}

\usepackage[accepted]{icml2019}

\icmltitlerunning{Neural MMO}

\begin{document}

\twocolumn[
\icmltitle{Neural MMO: A Massively Multiagent Game Environment \\ for Training and Evaluating Intelligent Agents}



\icmlsetsymbol{equal}{*}

\begin{icmlauthorlist}
\icmlauthor{Joseph Suarez}{}
\icmlauthor{Yilun Du}{}
\icmlauthor{Phillip Isola}{}
\icmlauthor{Igor Mordatch}{}
\end{icmlauthorlist}

\icmlcorrespondingauthor{}

\vskip 0.3in
]




\begin{abstract}
The emergence of complex life on Earth is often attributed to the arms race that ensued from a huge number of organisms all competing for finite resources. We present an artificial intelligence research environment, inspired by the human game genre of MMORPGs (Massively Multiplayer Online Role-Playing Games, a.k.a. MMOs), that aims to simulate this setting in microcosm. As with MMORPGs and the real world alike, our environment is persistent and supports a large and variable number of agents. Our environment is well suited to the study of large-scale multiagent interaction: it requires that agents learn robust combat and navigation policies in the presence of large populations attempting to do the same. Baseline experiments reveal that population size magnifies and incentivizes the development of skillful behaviors and results in agents that outcompete agents trained in smaller populations. We further show that the policies of agents with unshared weights naturally diverge to fill different niches in order to avoid competition.

\end{abstract}

\section{Introduction}
Life on Earth can be viewed as a massive multiagent competition. The cheetah evolves an aerodynamic profile in order to catch the gazelle, the gazelle develops springy legs to run even faster: species have evolved ever new capabilities in order to outcompete their adversaries. The success of biological evolution has inspired many attempts at creating ``artificial life" in silico. 

In recent years, the field of deep reinforcement learning (RL) has embraced a related approach: train agents by having them compete in simulated games \citep{silver2016mastering, OpenAI_dota, jaderberg2018human}. Such games are immediately interpretable and provide easy metrics derived from the game's ``score" and win conditions. However, popular game benchmarks typically define a narrow, episodic task with a small fixed number of players. In contrast, life on Earth involves a persistent environment, an unbounded number of players, and a seeming ``open-endedness", where ever new and more complex species emerge over time, with no end in sight~\citep{openendedness2017}.

\begin{figure*}
\begin{center}
\resizebox{0.97 \linewidth}{!}{
\includegraphics{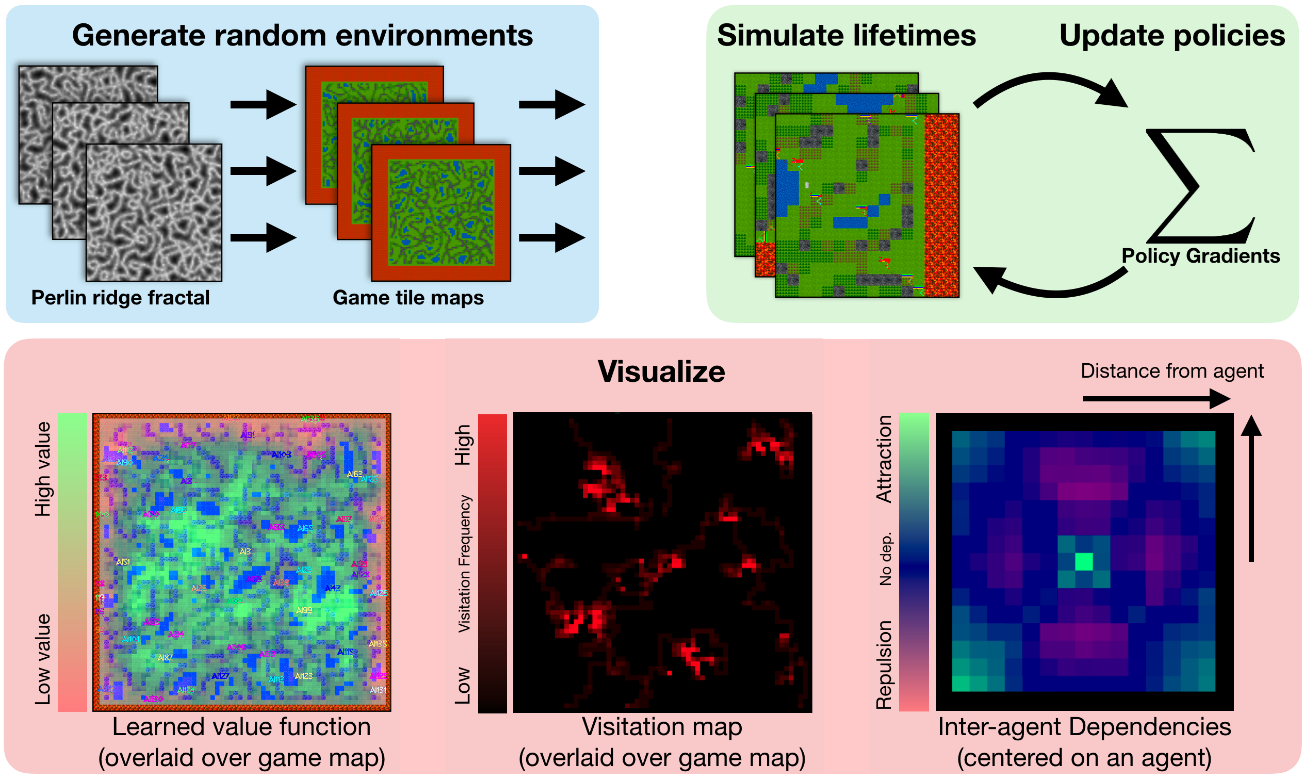}
}
\end{center}
\vspace{-4mm}
\caption{Our Neural MMO platform provides a procedural environment generator and visualization tools for value functions, map tile visitation distribution, and agent-agent dependencies of learned policies. Baselines are trained with policy gradients over 100 worlds.}
\label{fig:header}
\vspace{-1mm}
\end{figure*}

Our aim is to develop a simulation platform (see Figure \ref{fig:header}) that captures important properties of life on Earth, while also borrowing from the interpretability and abstractions of human-designed games. To this end, we turn to the game genre of Massively Multiplayer Online Role-Playing Games (MMORPGs, or MMOs for short). These games involve a large, variable number of players competing to survive and prosper in persistent and far-flung environments. Our platform simulates a ``Neural MMO" -- an MMO in which each agent is a neural net that learns to survive using RL.

We demonstrate the capabilities of this platform through a series of experiments that investigate emergent complexity as a function of the number of agents and species that compete in the simulation. We find that large populations act as competitive pressure that encourages exploration of the environment and the development of skillful behavior. In addition, we find that when agents are organized into species (share policy parameters), each species naturally diverges from the others to occupy its own behavioral niche. Upon publication, we will opensource the platform in full.

\section{Background and Related Work}

\paragraph{Artificial Life and Multiagent Reinforcement Learning}
Research in ``Artificial life" aims to model evolution and natural selection in biological life; \citep{langton1997artificial, ficici1998challenges}. Such projects often consider open-ended skill learning ~\citep{yaeger1994computational} and general morphology evolution ~\citep{sims1994evolving2} as primary objectives. Similar problems have recently resurfaced within multiagent reinforcement learning where the continual co-adaptation of agents can introduce additional nonstationarity that is not present in single agent environments. While there have been multiple attempts to formalize the surrounding theory \citep{hernandez2011more, strannegard2018}, we primarily consider environment-driven works. These typically consider either complex tasks with 2-10 agents \citep{bansal2017emergent, OpenAI_dota, jaderberg2018human} or much simpler environments with tens to upwards of a million agents ~\citep{lowe2017multi, mordatch2017emergence, bansal2017emergent, lanctot2017unified, yang2018mean, zheng2017magent, jaderberg2018human}. Most such works further focus on learning a specific dynamic, such as predator-prey \citet{yang2018study} or are more concerned with the study than the learning of behavior, and use hard-coded rewards \citet{zheng2017magent}. In contrast, our work focuses on large agent populations in complex environments.

\paragraph{Game Platforms for Intelligent Agents}
The Arcade Learning Environment (ALE)~\citep{bellemare2013arcade} and Gym Retro~\citep{nichol2018gotta} provide 1000+ limited scope arcade games most often used to test individual research ideas or generality across many games. Better performance at a large random subset of games is a reasonable metric of quality. However, recent results have brought into question the overall complexity each individual environment ~\citep{cuccu2018playing}, and strong performance in such tasks is not particularly difficult for humans.

More recent work has demonstrated success on multiplayer games including Go~\citep{silver2016mastering}, the Multiplayer Online Battle Arena (MOBA) game DOTA2~\citep{OpenAI_dota}, and Quake 3 Capture the Flag ~\citep{jaderberg2018human}. Each of these projects has advanced our understanding of a class of algorithms. However, these games are limited to 2-12 players, are episodic, with game rounds on the order of an hour, lack persistence, and lack the game mechanics supporting large persistent populations -- there is still a large gap in environment complexity compared to the real world.

\paragraph{Role-playing games (RPGs)}
such as Pokemon and Final Fantasy, are in-depth experiences designed to engage human players for hundreds of hours of persistent gameplay. Like the real world, problems in RPGs have many valid solutions and choices have long term consequences.

\paragraph{MMORPGs}
are the (massively) multiplayer analogs to RPGs. They are typically run across several persistent servers, each of which contains a copy of the environment and supports hundreds to millions of concurrent players. Good MMOs require increasingly clever, team-driven usage of the game systems: players attain the complex skills and knowledge required for the hardest challenges only through a curriculum of content spanning hundreds of hours of gameplay. Such a curriculum is present in many game genres, but only MMOs contextualize it within persistent social and economic structures approaching the scale of the real world.

\begin{figure*}[ht!]
\begin{center}
\resizebox{0.96\linewidth}{!}{
\includegraphics{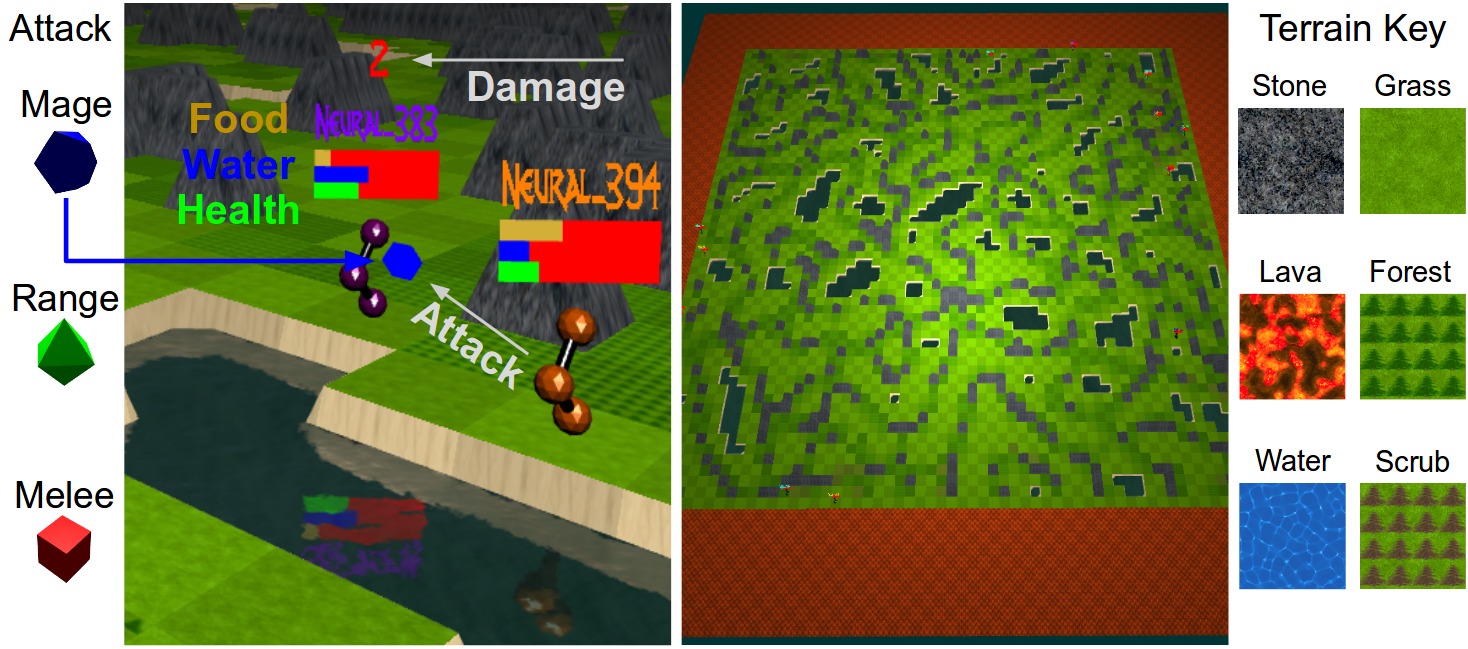}
}
\end{center}
\vspace{-4mm}
\caption{Our platform includes an animated 3D client and a toolbox used to produce the visuals in this work. Agents compete for food and water while engaging in strategic combat. See the Neural MMO section for a brief overview and the Supplement for full details.}\label{fig:cent}
\label{fig:environment}
\vspace{-2mm}
\end{figure*}

\section{Neural MMO}

We present a persistent and massively multiagent environment that defines foraging and combat systems over procedurally generated maps. The Supplement provides full environment details and Figure \ref{fig:environment} shows a snapshot. The core features are support for a large and variable number of agents, procedural generation of tile-based terrain, a food and water foraging system, a strategic combat system, and inbuilt visualization tools for analyzing learned policies

Agents (players) may join any of several servers (environment instances). Each server contains an automatically generated tile-based environment of configurable size. Some tiles, such as food-bearing forest tiles and grass tiles, are traversable. Others, such as water and solid stone, are not. Upon joining a server, agents spawn at a random location along the edges of the environment. In order remain healthy (maintain their health statistic), agents must obtain food and water -- they die upon reaching reaching 0 health. At each \textbf{server tick (time step)}, agents may move one tile and make an attack. Stepping on a forest tile or next to a water tile refills a portion of the agent's food or water supply, respectively. However, forest tiles have a limited supply of food; once exhausted, food has a 2.5 percent chance to regenerate each tick. This means that agents must compete for food tiles while periodically refilling their water supply from infinite water tiles. They may attack each other using any of three attack options, each with different damage values and tradeoffs. Precise foraging and combat mechanics are detailed in the Supplement.

Agents observe local game state and decide on an action each game tick. The environment does not make any further assumptions on the source of that decision, be it a neural network or a hardcoded algorithm. We have tested the environment with up to 100 million agent trajectories (lifetimes) on ~100 cores in ~1 week. Real and virtual worlds alike are open-ended tasks where complexity arises with little direction. Our environment is designed as such. Instead of rewarding agents for achieving particular objectives optimize only for survival time: they receive reward $r_t=1$ for each time step alive. Competition for finite resources mandates that agents must learn intelligent strategies for gathering food and water in order to survive.

One purpose of the platform is to discover \textit{game mechanics} that support complex behavior and \textit{agent populations} that can learn to make use of them. In human MMOs, \textit{developers aim to create balanced mechanics while players aim to maximize their skill in utilizing them}. The initial configurations of our systems are the results of several iterations of balancing, but are by no means fixed: every numeric parameter presented is editable within a simple configuration file.

\begin{figure*} 
\centering
\begin{tabular}{cc}
    \subfloat{\includegraphics[width=0.4\linewidth]{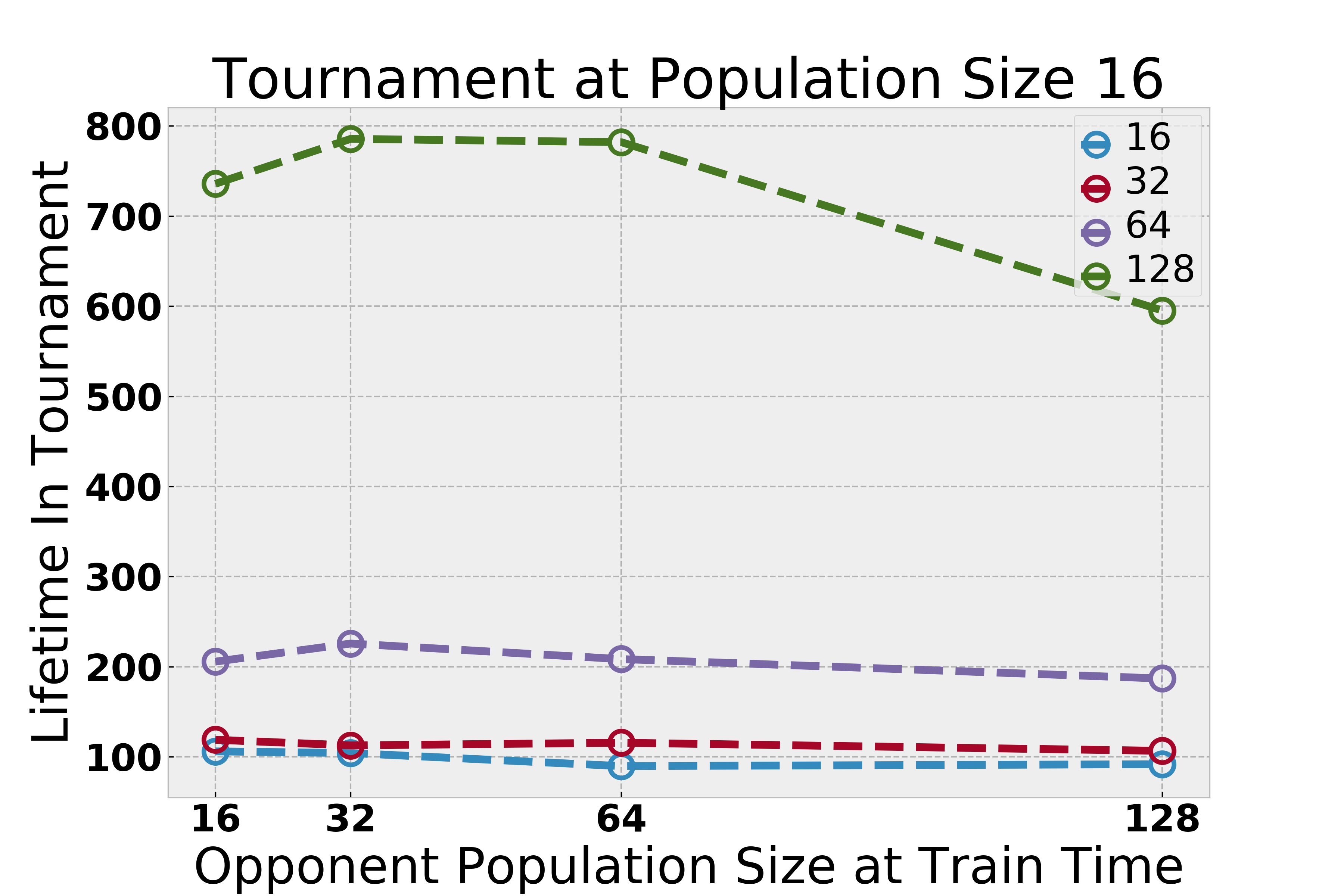}}
    \subfloat{\includegraphics[width=0.4\linewidth]{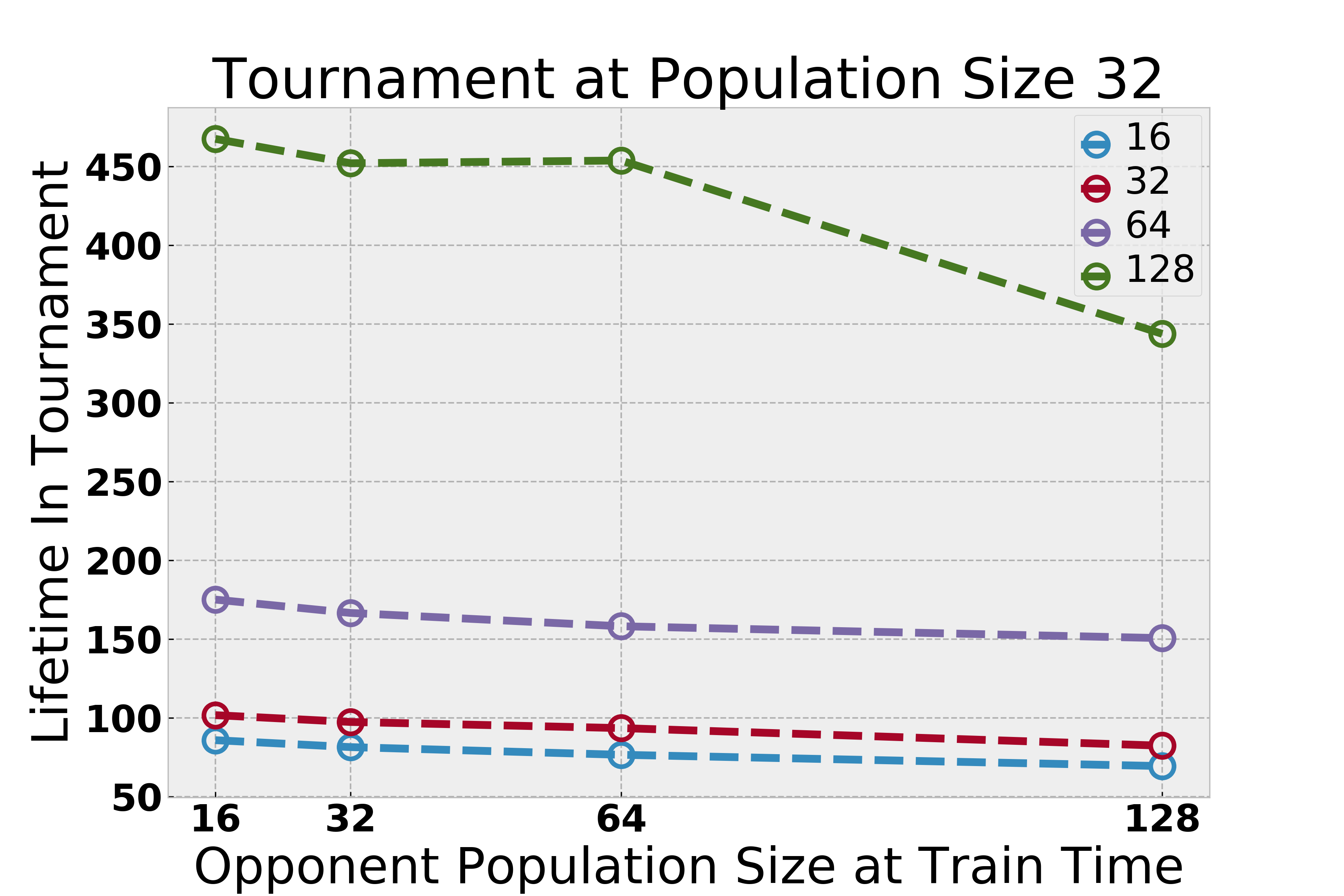}} \\
    \subfloat{\includegraphics[width=0.4\linewidth]{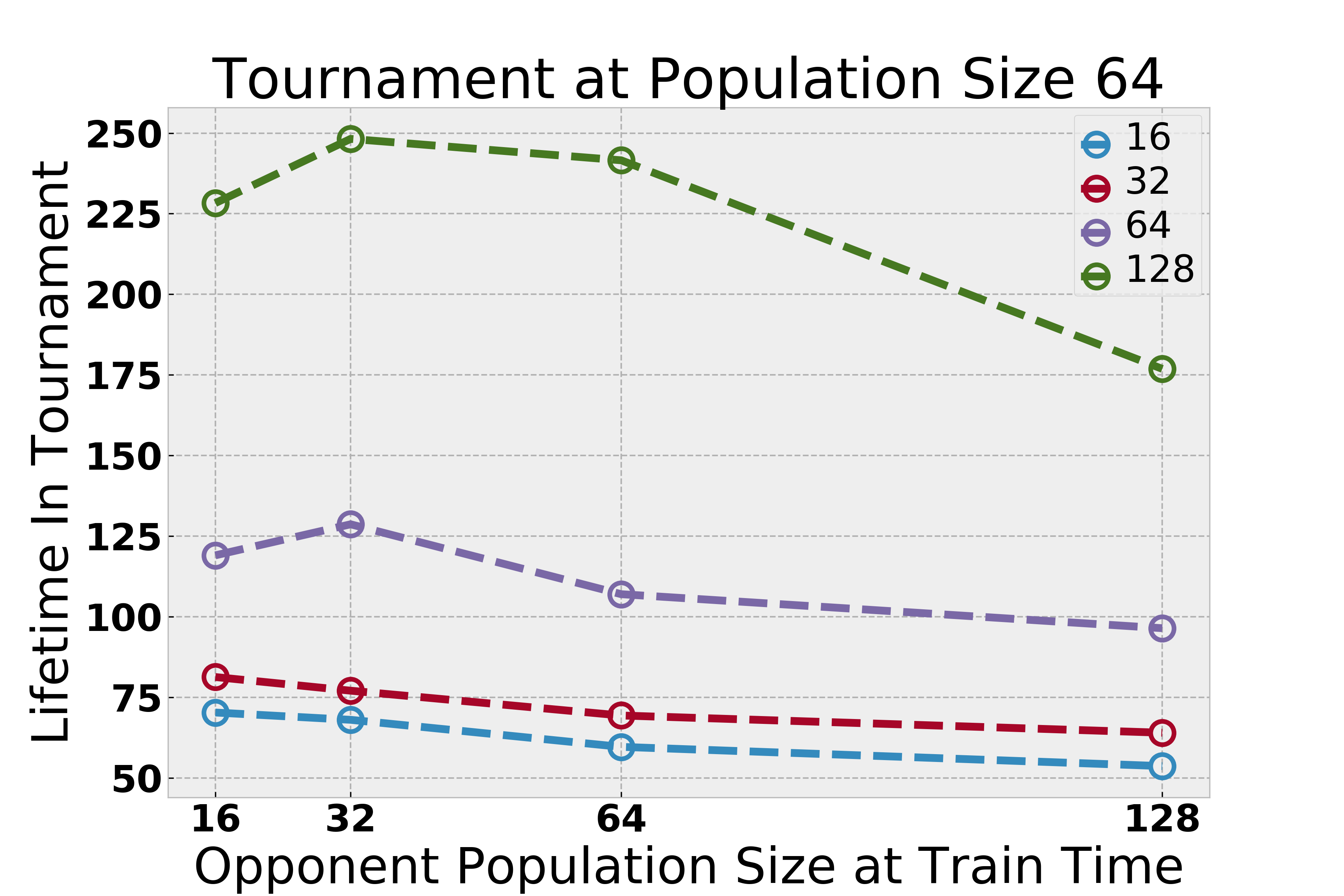}}
    \subfloat{\includegraphics[width=0.4\linewidth]{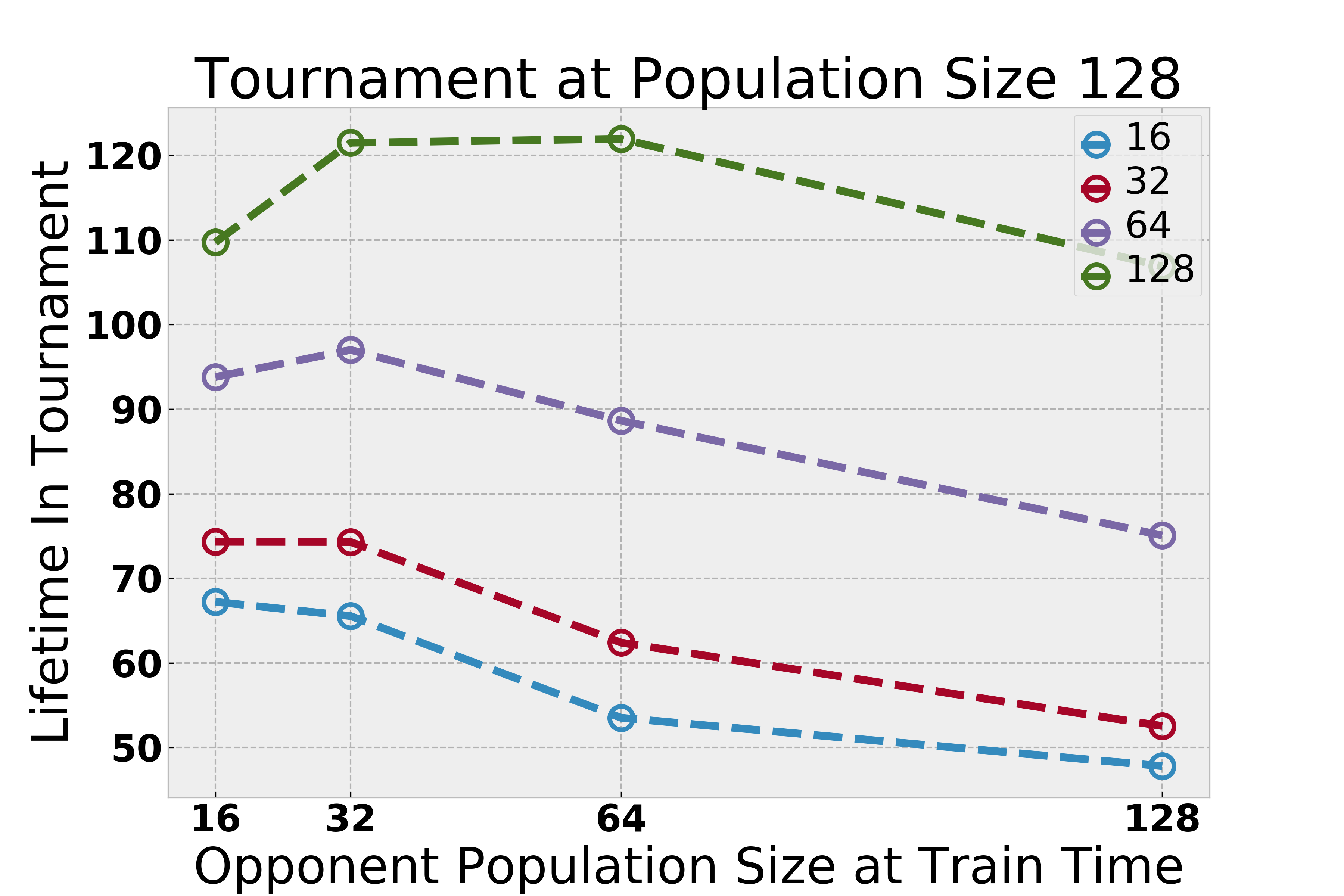}}
\end{tabular}
\vspace{-3mm}
\caption{Maximum population size at train time varies in (16, 32, 64, 128). At test time, we merge the populations learned in pairs of experiments and evaluate lifetimes at a fixed population size. Agents trained in larger populations always perform better.}
\label{fig:tournament}
\end{figure*}

\begin{figure*}
\begin{center}
\includegraphics[width=0.90\linewidth]{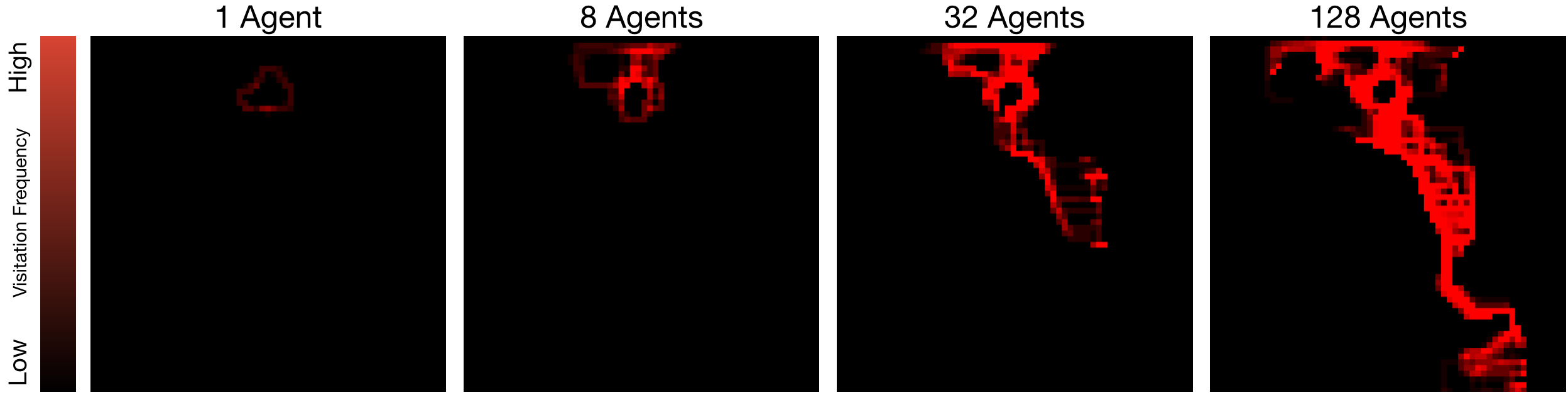}
\vspace{-3mm}
\caption{Population size magnifies exploration: agents spread out to avoid competition. }\label{fig:fent}
\label{fig:explore}
\end{center}
\end{figure*}

\begin{figure*}
\begin{center}
\includegraphics[width=0.90\linewidth]{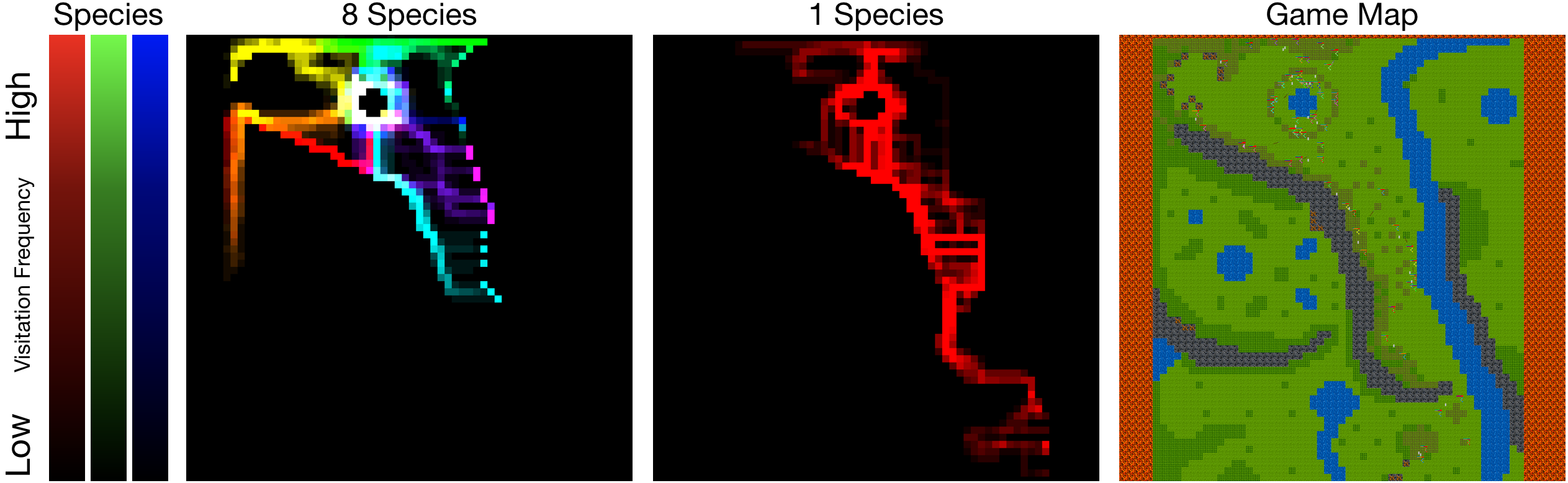}
\vspace{-3mm}
\caption{Populations count (number of species) magnifies niche formation. Visitation maps are overlaid over the game map; different colors correspond to different species. Training a single population tends to produce a single deep exploration path. Training eight populations results in many shallower paths: populations spread out to avoid competition among species.}
\label{fig:niche}
\end{center}
\end{figure*}

\section{Architecture and Training}
Agents are controlled by policies parameterized by neural networks. Agents make observations $o_t$ of the game state $s_t$ and follow a policy $\pi(o_t) \to a_t$ in order to make actions $a_t$. We maximize a return function $R$ over trajectory $\tau = (o_t, a_t, r_t, ..., o_T, a_T, r_T)$. This is a discounted sum of survival rewards: $R(\tau) = \sum_t^T \gamma^tr_t$ where $\gamma = 0.99$, $T$ is the time at death and the survival reward $r_t$ equals 1, as motivated previously. The policy $\pi$ may be different for each agent or shared. Algorithm \ref{alg:pipeline} shows high level training logic. The Supplement details the tile-based game state $s_t$ and hyperparameters (Table 1). 
\clearpage

\begin{algorithm}[tb]
   \caption{Neural MMO logic for one game tick. See Experiments (Technical details) for spawning logic. The algorithm below makes two omissions for simplicity. First, we use multiple policies and sample a policy $\pi \sim {\pi_1, \ldots, \pi_N}$ from the set of all policies when spawning a new agent. Second, instead of performing a policy gradient update every game tick, we maintain experience buffers from each environment and perform an update once all buffers are full.}
   \label{alg:pipeline}
\begin{algorithmic}
   \FOR{\textbf{each} environment server}
      \IF{number of agents alive  $<$ spawn cap}
         \STATE spawn an agent
      \ENDIF
      \FOR{\textbf{each} agent}
         \STATE i $\leftarrow$ population index of the agent
         \STATE Make observation $o_t$, decide action $\pi_i(o_t) \to a_t$
         \STATE Environment processes $a_t$, computes $r_t$, and updates agent health, food, etc.
         \IF{agent is dead}
            \STATE remove agent
         \ENDIF
      \ENDFOR
      \STATE Update environment state $s_{t+1} \rightarrow f(s_t, a_t)$
   \ENDFOR
   \STATE Perform a policy gradient update on policies $\pi \sim {\pi_1, \ldots, \pi_N}$ using $o_t$, $a_t$, $r_t$ from all agents across all environment servers
\end{algorithmic}
\end{algorithm}
\vspace{-3mm}
\textbf{Input} We set the observation state $o_t$ equal to the crop of tiles within a fixed $L_1$ distance of the current agent. This includes tile terrain types and the select properties (such as health, food, water, and position) of occupying agents. Our choice of $o_t$ is an equivalent representation of what a human sees on the screen, but our environment supports other choices as well. Note that computing observations does not require rendering.

\textbf{Output} Agents output action choices $a_t$ for the next time step (game tick). Actions consist of one movement and one attack. Movement options are: North, South, East, West, and Pass (no movement). Attack options are labeled: Melee, Range, and Mage, with each attack option applying a specific preset amount of damage at a preset effective distance. The environment will attempt to execute both actions. Invalid actions, (\textit{e.g.} moving into stone), are ignored.

Our policy architecture preprocesses the local environment by embedding it and flattening it into a single fixed length vector. We then apply a linear layer followed by linear output heads for movement and attack decisions. New types of action choices can be included by adding additional heads. We also train a value function to estimate the discounted return. As agents receive only a stream of reward 1, this is equal to a discounted estimate of the agent's time until death. We use a value function baselines policy gradient loss and optimize with Adam. It was possible to obtain good performance without discounting, but training was less stable. We provide full details in the supplements.

\section{Experiments}

We present an initial series of experiments using our platform to explore multiagent interactions in large populations. We find that agent competence scales with population size. In particular, increasing the maximum number of concurrent players ($N_{ent}$) magnifies exploration and increasing the maximum number of populations with unshared weights ($N_{pop}$) magnifies niche formation. Agents policies are sampled uniformly from a number of ``populations" $\pi \sim {\pi_1, \ldots, \pi_N} $. Agents in different populations have the same architecture but do not share weights.

\textbf{Technical details} We run each experiment using 100 worlds. We define a constant $C$ over the set of worlds $W$. For each world $w\in W$, we uniformly sample a $c \in (1, 2, ... C)$. We define "spawn cap" such that if world $w$ has a spawn cap $c$, the number of agents in $w$ cannot exceed $c$. In each world $w$, one agent is spawned per game tick provided that doing so would exceed the spawn cap $c$ of $w$. To match standard MMOs, we would fix $N_{ent}$ = $N_{pop}$ (humans are independent networks with unshared weights). However, this incurs sample complexity proportional to number of populations. We therefore share parameters across groups of up to 16 agents for efficiency.

\subsection{Server Merge Tournaments}

We perform four experiments to evaluate the effects on foraging performance of training with larger populations and with a greater number of populations. For each experiment, we fix $N_{pop} \in (1, 2, 4, 8)$ and a spawn cap (the maximum number of concurrent agents) $c=16 \times N_{pop}$, such that $c \in (16, 32, 64, 128)$. We train for a fixed number of trajectories per population.

Evaluating the influence of these variables is nontrivial. The task difficulty is highly dependent on the size and competence of populations in the environment: mean agent lifetime is not comparable across experiments. Furthermore, there is no standard procedure among MMOs for evaluating relative player competence across multiple servers. However, MMO servers sometimes undergo merges whereby the player bases from multiple servers are placed within a single server. As such, we propose tournament style evaluation in order to directly compare policies learned in different experiment settings. Tournaments are formed by simply concatenating the player bases of each experiment. Figure \ref{fig:tournament} shows results: we vary the maximum number of agents at test time and find that agents trained in larger settings consistently outperform agents trained in smaller settings.
\clearpage
We observe more interesting policies once we introduce the combat module as an additional learnable mode of variation on top of foraging. With combat, agent actions become strongly coupled with the states of other agents. As a sanity check, we also confirm that all of the populations trained with combat handily outperform all of the populations trained with only foraging, when these populations compete in a tournament with combat enabled.

To better understand theses results, we decouple our analysis into two modes of variability: maximum number of concurrent players ($N_{ent}$) and maximum number of populations with unshared weights ($N_{pop}$). This allows us to examine the effects of each factor independently. In order to isolate the effects of environment randomization, which also encourages exploration, we perform these experiments on a fixed map. Isolating the effects of these variables produces more immediately obvious results, discussed in the following two subsections:

\subsection{$N_{ent}$: Multiagent Magnifies Exploration}

In the natural world, competition between animals can incentivize them to spread out in order to avoid conflict. We observe that overall exploration (map coverage) increases as the number of concurrent agents increases (see Figure \ref{fig:explore}; the map used is shown in Figure \ref{fig:niche}). Agents learn to explore only because the presence of other agents provides a natural incentive for doing so.

\subsection{$N_{pop}$: Multiagent Magnifies Niche Formation}

We find that, given a sufficiently large and resource-rich environment, different populations of agents tend to separate to avoid competing with other populations. Both MMOs and the real world often reward masters of a single craft more than jacks of all trades. From Figure \ref{fig:niche}, specialization to particular regions of the map increases as number of populations increases. This suggests that the presence of other populations force agents to discover a single advantageous skill or trick. That is, increasing the number of populations results in diversification to separable regions of the map. As entities cannot out-compete other agents of their own population (i.e. agent's with whom they share weights), they tend to seek areas of the map that contain enough resources to sustain their population.

\subsection{Environment Randomized Exploration}
The trend of increasing exploration with increasing entity number is clear when training on a single map as seen in Figure \ref{fig:explore}, \ref{fig:niche}, but it is more subtle with environment randomization. From Figure \ref{fig:rand}, all population sizes explore adequately. It is likely that ``exploration" as defined by map coverage is not as difficult a problem, in our environment, as developing robust policies. As demonstrated by the Tournament experiments, smaller populations learn brittle policies that do not generalize to scenarios with more competitive pressure--even against a similar number of agents.

\subsection{Agent-Agent Dependencies}
We visualize agent-agent dependencies in Figure \ref{fig:deps}. We fix an agent at the center of a hypothetical map crop. For each position visible to that agent, 
we show what the value function would be if there were a second agent at that position.
We find that agents learn policies dependent on those of other agents, in both the foraging and combat environments.

\section{Discussion}
\subsection{Multiagent competition is a curriculum magnifier}
Not all games are created equal. Some produce more complex and engaging play than others. It is unreasonable to expect pure multiagent competition to produce diverse and interesting behavior if the environment does not support it. This is because \textbf{multiagent competition is a curriculum \textit{magnifier}, not a curriculum in and of itself}. The initial conditions for formation of intelligent life are of paramount importance. Jungle climates produce more biodiversity than deserts. Deserts produce more biodiversity than the tallest mountain peaks. To current knowledge, Earth is the only planet to produce life at all. The same holds true in simulation: human MMOs mirror this phenomenon. Those most successful garner large and dedicated player bases and develop into complex ecosystems. The multiagent setting is interesting because learning is responsive to the competitive and collaborative pressures of other learning agents--but the environment must support and facilitate such pressures in order for multiagent interaction to drive complexity.

There is room for debate as to the theoretical simplest possible seed environment required to produce complexity on par with that of the real world. However, this is not our objective. We have chosen to model our environment after MMOs, even though they may be more complicated than the minimum required environment class, because they are known to support the types of interactions we are interested in while maintaining engineering and implementation feasibility. This is not true of any other class environments we are aware of: exact physical simulations are computationally infeasible, and previously studied genres of human games lack crucial elements of complexity (see Background). While some may see our efforts as cherrypicking environment design, we believe this is precisely the objective: the primary goal of game development is to create complex and engaging play at the level of human intelligence. The player base then uses these design decisions to create strategies far beyond the imagination of the developers.


\begin{figure*}
\includegraphics[width=\linewidth]{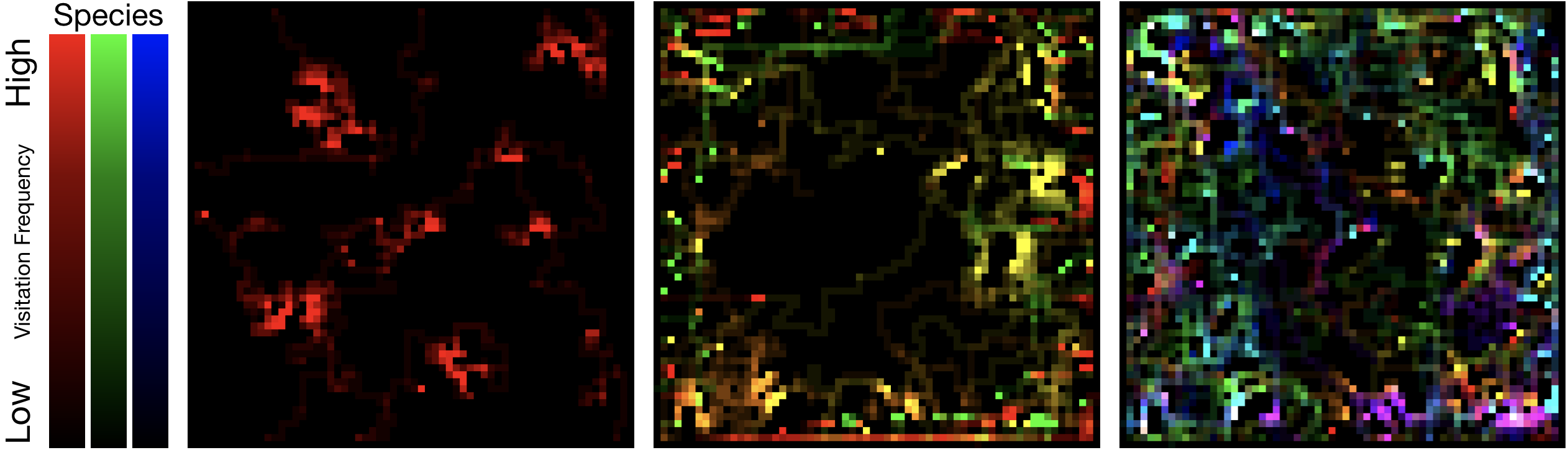}
\caption{Exploration maps in the environment randomized settings. From left to right: population size 8, 32, 128. All populations explore well, but larger populations with more species develop robust and efficient policies that do better in tournaments.}
\label{fig:rand}

\vspace{0.5cm}
\includegraphics[width=\linewidth]{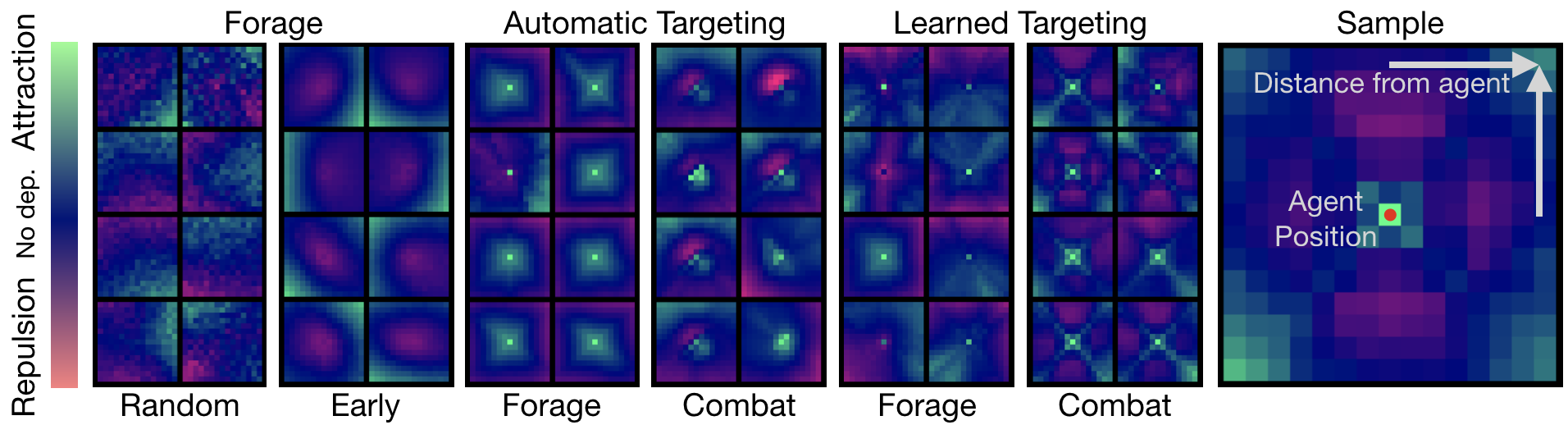}
\caption{Agents learn to depend on other agents. Each square map shows the response of an agent of a particular species, located at the square's center, to the presence of agents at any tile around it. Random: dependence map of random policies. Early: "bulls eye" avoidance maps learned after only a few minutes of training. Additional maps correspond to foraging and combat policies learned with automatic targeting (as in tournament results) and learned targeting (experimental, discussed in Additional Insights). In the learned targeting setting, agents begin to fixate on the presence of other agents within combat range, as denoted by the central square patterns.}
\label{fig:deps}

\vspace{0.5cm}
\includegraphics[width=\linewidth]{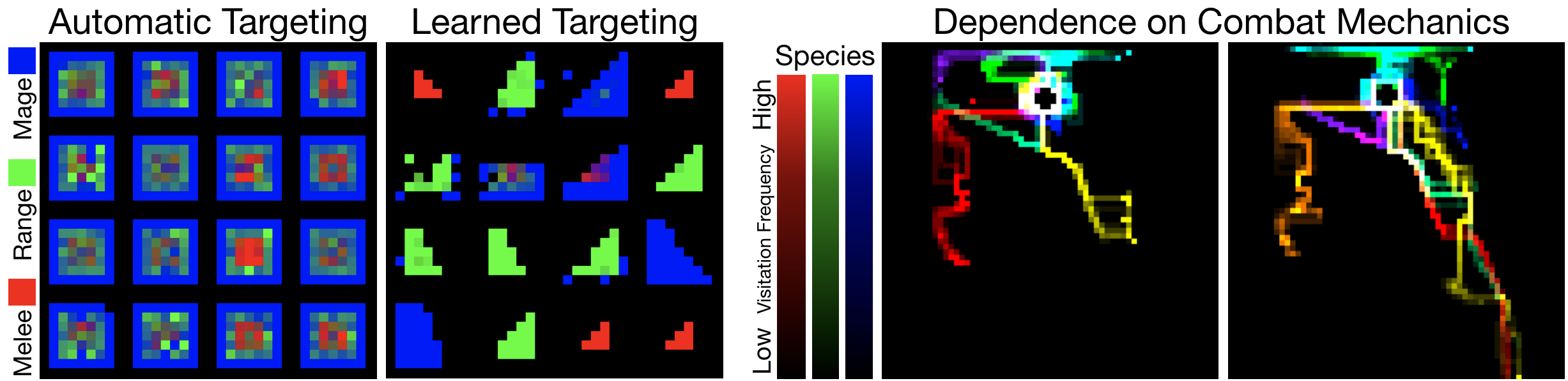}
\caption{Attack maps and niche formation quirks. Left: combat maps from automatic and learned targeting. The left two columns in each figure are random. Agents with automatic targeting learn to make effective use of melee combat (denoted by higher red density). Right: noisy niche formation maps learned in different combat settings with mixed incentives to engage in combat.}
\label{fig:misc}
\end{figure*}
\clearpage

\subsection{Additional Insights}
We briefly detail several miscellaneous points of interest in Figure \ref{fig:misc}. First, we visualize learned attack patterns of agents. Each time an agent attacks, we splat the attack type to the screen. There are a few valid strategies as per the environment. Melee is intentionally overpowered, as a sanity check. This cautions agents to keep their distance, as the first to strike wins. We find that this behavior is learned from observation of the policies learned in Figure \ref{fig:misc}.

Second, a note on tournaments. We equate number of trajectories trained upon as a fairest possible metric of training progress. We experimented with normalizing batch size but found that larger batch size always leads to more stable performance. Batch size is held constant, but experience is split among species. This means that experiments with more species have smaller effective batch size: larger populations outperform smaller populations even though the latter are easier to train.

Finally, a quick note on niche formation. Obtaining clean visuals is dependent on having an environment where interaction with other agents is unfavorable. While we ensure this is the case for our exploration metrics, niche formation may also occur elsewhere, such as in the space of effective combat policies. For this reason, we expect our environment to be well suited to methods that encourage sample diversity such as population-based training~\citep{jaderberg2017population}.

\section{Future Work}
Our final set of experiments prescribes targeting to the agent with lowest health. Learned targeting was not required to produce compelling policies: agents instead learn effective attack style selection, strafing and engaging opportunistically at the edge of their attack radius. Another possible experiment is to jointly learn attack style selection and targeting. This would require an attentional mechanism to handle the variable number of visible targets. We performed only preliminary experiments with such an architecture, but we still mention them here because even noisy learned targeting policies significantly alter agent-agent dependence maps. As shown in Figure \ref{fig:deps}, the small square shaped regions of high value at the center of the dependency maps correspond to the ranges of different attack styles. These appear responsive to the current combat policies of other learning agents. We believe that the learned targeting setting is likely to useful for investigating the effects of concurrent learning in large populations.

\section{Conclusion}
We have presented a neural MMO as a research platform for multiagent learning. Our environment supports a large number of concurrent agents, inbuilt map randomization, and detailed foraging and combat systems. The included baseline experiments demonstrate our platform's capacity for research purposes. We find that population size magnifies exploration in our setting, and the number of distinct species magnifies niche formation. It is our hope that our environment provides an effective venue for multiagent experiments, including studies of niche formation, emergent cooperation, and coevolution. The entire platform will be open sourced, including a performant 3D client and research visualization toolbox. Full technical details of the platform are available in the Supplement.

\section*{Acknowledgements}
This research was undertaken in fulfillment of an internship at OpenAI. Thank you to Clare Zhu for substantial contributions to the 3D client code.

\bibliography{example_paper}
\bibliographystyle{icml2019}



\twocolumn[
\icmltitle{Neural MMO Supplement}
]

\begin{figure}
\begin{center}
\resizebox{1.0\linewidth}{!}{
\includegraphics{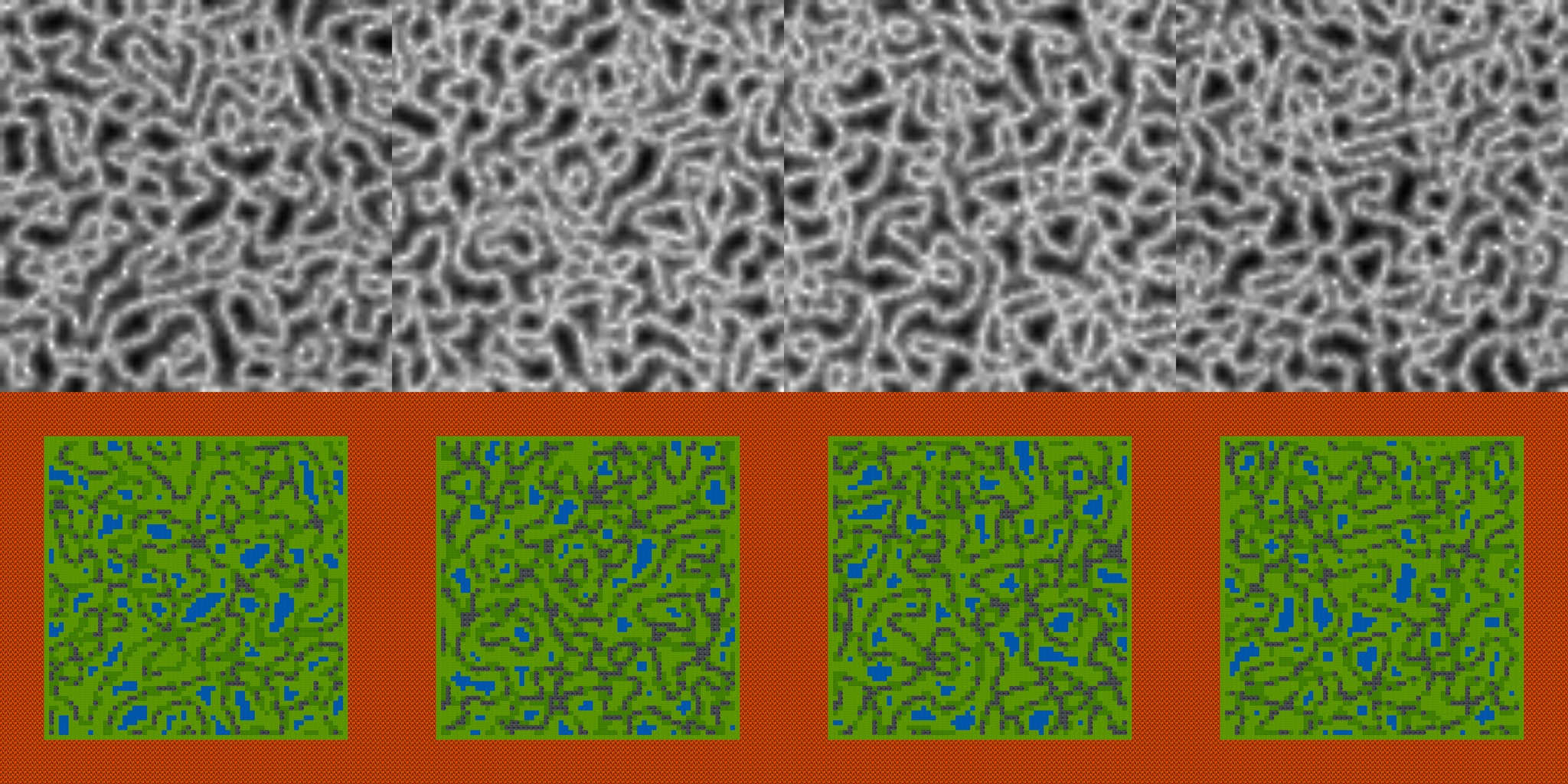}
}
\end{center}
\vspace{-5mm}
\caption{Procedural 80X80 game maps}
\label{fig:procedural}
\vspace{-5mm}
\end{figure}

\section*{Environment}
The environment state is represented by a grid of tiles. We generate the game map by thresholding a Perlin \citep{Perlin:1985:IS:325165.325247} ridge fractal, as shown in Figure \ref{fig:procedural}. Each tile has a particular assigned material with various properties, but it also maintains a set of references to all occupying entities. When agents observe their local environment, they are handed a crop of all visible game tiles, including all visible properties of the tile material and all visible properties of occupying agents. All parameters in the following subsystems are configurable; we provide only sane defaults obtained via multiple iterations of balancing.

\subsection*{Tiles}
We adopt a tile based game state, which is common among MMOs. This design choice is computationally efficient for neural agents and can be made natural for human players via animation smoothing. When there is no need to render the game client, as in during training or test time statistical tests, the environment can be run with no limit on server tick rate. Game tiles are as follows:

\begin{itemize}
    \setlength\itemsep{0pt}
    \item \textbf{Grass:} Passable tile with no special properties
    \item \textbf{Forest:} Passable tile containing food. Upon moving into a food tile, the agent gains 5 food and the tile decays into a scrub. 
    \item \textbf{Scrub:} Passable tile that has a 2.5 percent probability to regenerate into a forest tile on each subsequent tick
    \item \textbf{Stone:} Impassible tile with no special properties
    \item \textbf{Water:} Passable tile containing water. Upon moving adjacent to a water tile, the agent gains 5 water.
    \item \textbf{Lava:} Passable tile that kills the agent upon contact
\end{itemize}

\begin{figure}
\begin{center}
\resizebox{1.0\linewidth}{!}{
\includegraphics{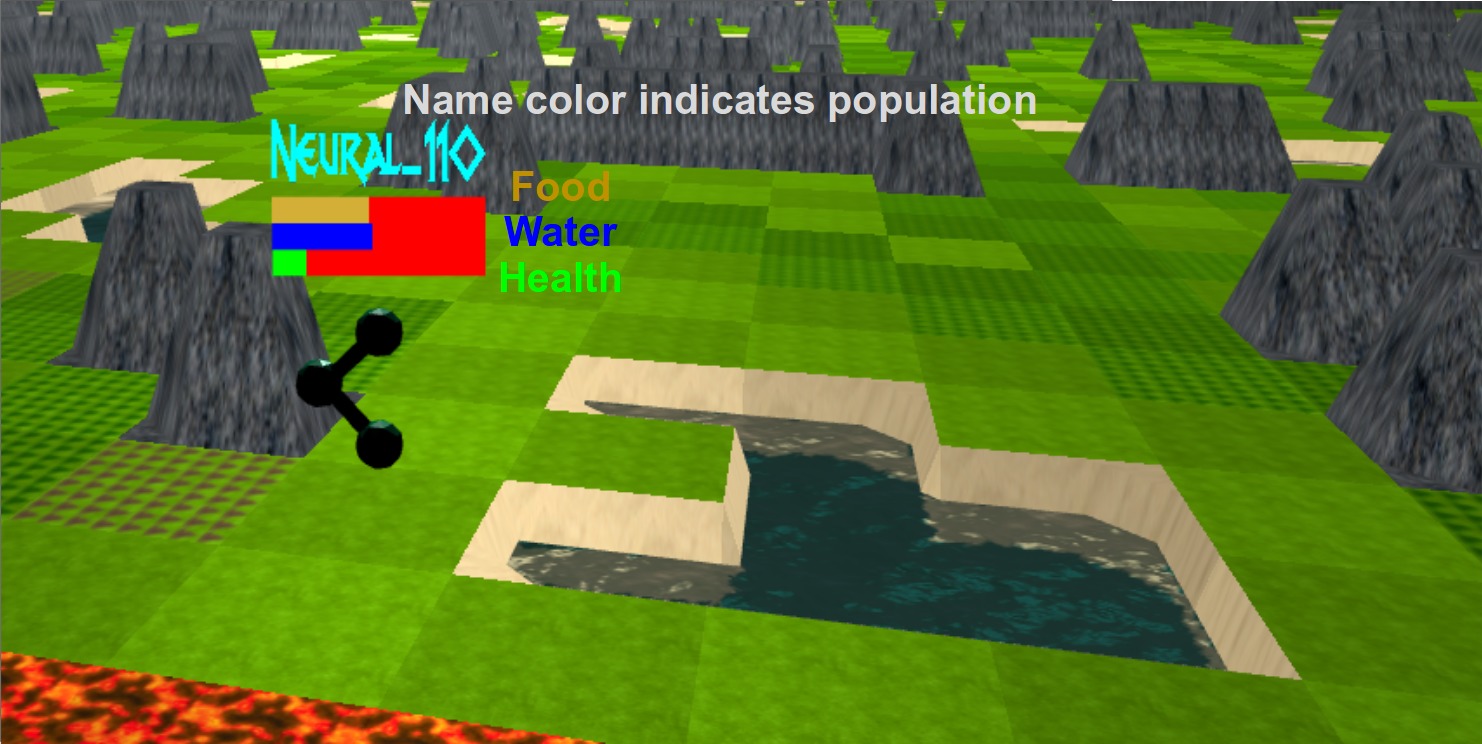}
}
\end{center}
\vspace{-5mm}
\caption{Example Agent}
\label{fig:agent}
\vspace{-5mm}
\end{figure}

\section*{Agents}
\paragraph{Input:} On each game tick, agents (Figure \ref{fig:agent}) observe a 15x15 square crop of surrounding game tiles and all occupying agents. We extract the following observable properties: 

\textbf{Per-tile properties:}
\begin{itemize}
    \setlength\itemsep{0pt}
    \item \textbf{Material}: an index corresponding to the tile type
    \item \textbf{nEnts}: The number of occupying entities. This is technically learnable from the list of agents, but this may not be true for all architectures. We include it for convenience here, but may deprecate it in the future.
\end{itemize}

\textbf{Per-agent properties:}
\begin{itemize}
    \setlength\itemsep{0pt}
    \item \textbf{Lifetime}: Number of game ticks alive thus far
    \item \textbf{Health}: Agents die at 0 health (hp)
    \item \textbf{Food}: Agents begin taking damage at 0 food or water
    \item \textbf{Water}: Agents begin taking damage at 0 food or water
    \item \textbf{Position}: Row and column of the agent
    \item \textbf{Position Deltas}: Offsets from the agent to the observer
    \item \textbf{Damage}: Most recent amount of damage taken
    \item \textbf{Same Color}: Whether the agent is the same color (and thereby is in the same population) as the observer
    \item \textbf{Freeze}: Whether the agent is frozen in place as a result of having been hit by a mage attack
\end{itemize}

\paragraph{Output:} Agents submit one movement and one attack action request per server tick. The server ignores any actions that are not possible or permissible to fulfil, such as attacking an agent that is already dead or attempting to move into stone. \textit{Pass} corresponds to no movement.

\begin{tabular}{ l l l l l l }
  \textbf{Movement:} & \textbf{North} & \textbf{South} & \textbf{East} & \textbf{West} & \textbf{Pass} \\
  \textbf{Attack:} & \textbf{Melee} & \textbf{Range} & \textbf{Mage} \\
\end{tabular}

\begin{figure}
\begin{center}
\resizebox{1.0\linewidth}{!}{
\includegraphics{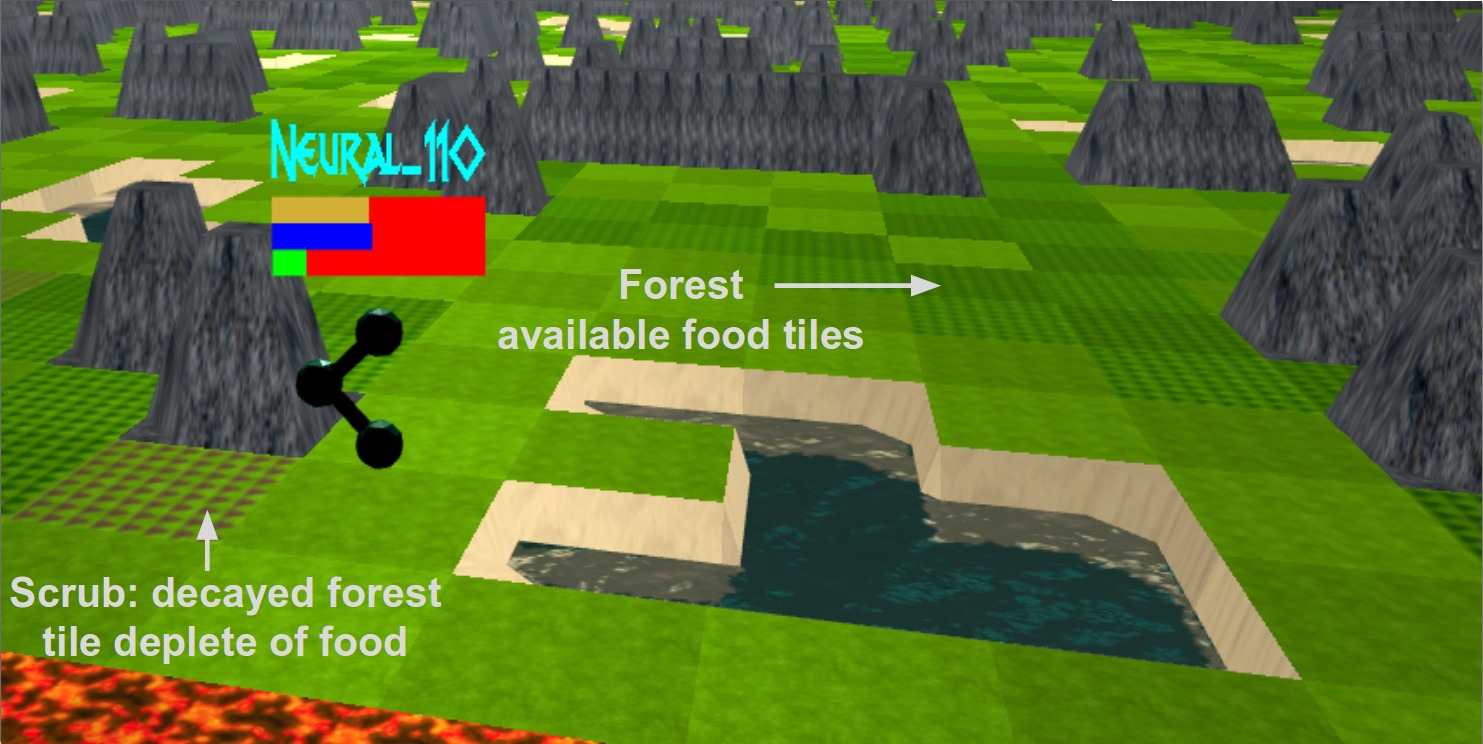}
}
\end{center}
\vspace{-5mm}
\caption{Example foraging behavior}
\label{fig:forage}
\vspace{-6mm}
\end{figure}
\section*{Foraging}
Foraging (Figure \ref{fig:forage}) implements gathering based survival:
\begin{itemize}
    \setlength\itemsep{0pt}
    \item \textbf{Food:} Agents begin with 32 food, decremented by 1 per tick. Agents may regain food by occupying forest tiles or by making use of the combat system.
    \item \textbf{Water:} Agents begin with 32 water, decremented by 1 per tick. Agents may regain water by occupying tiles adjacent to water or making use of the combat system. 
    \item \textbf{Health:} Agents begin with 10 health. If the agent hits 0 food, they lose 1 health per tick. If the agent hits 0 water, they lose 1 health per tick. These effects stack.
\end{itemize}

The limited availability of forest (food) tiles produces a carrying capacity. This incurs an arms race of exploration strategies: survival is trivial with a single agent, but it requires intelligent exploration in the presence of competing agents attempting to do the same.  

\section*{Combat}
\begin{figure}
\begin{center}
\resizebox{1.0\linewidth}{!}{
\includegraphics{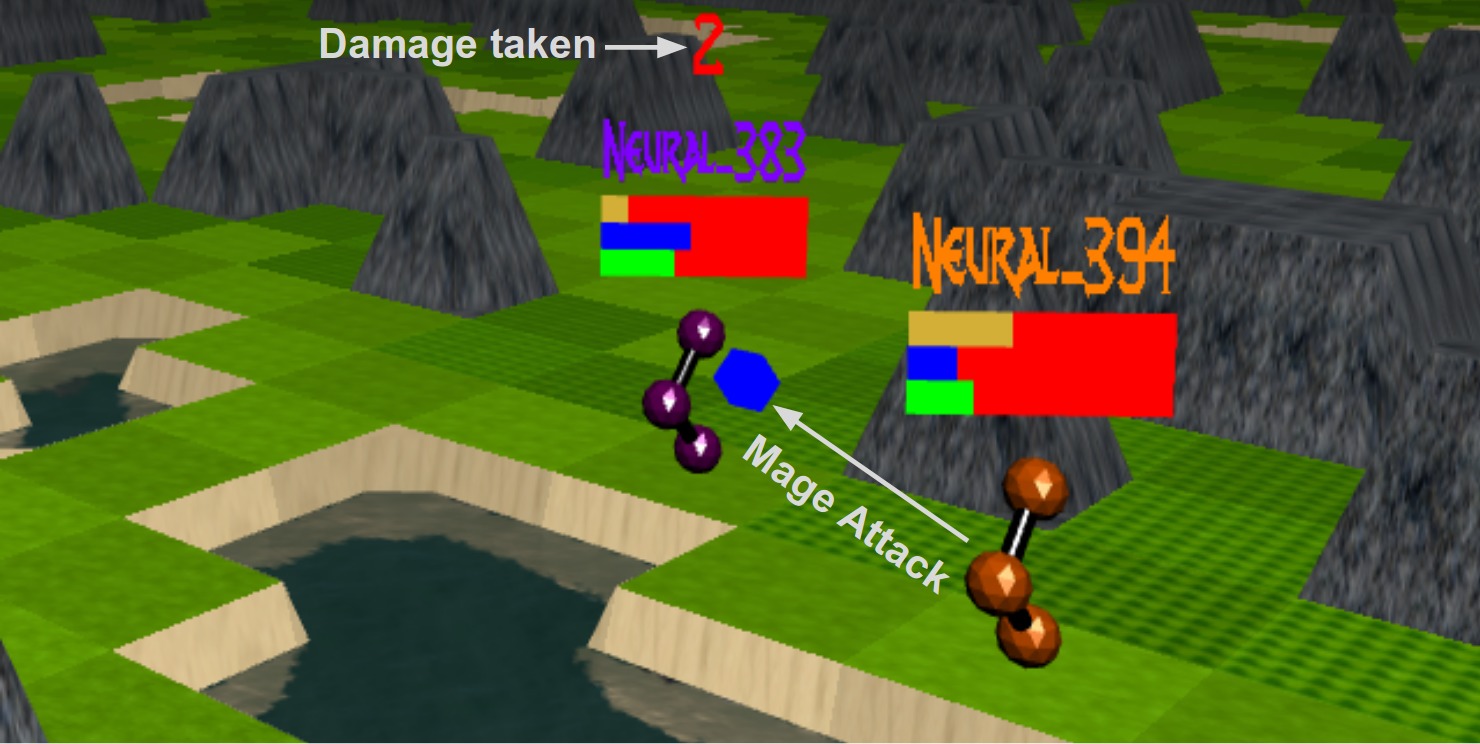}
}
\end{center}
\vspace{-5mm}
\caption{Example combat behavior}
\label{fig:combat}
\vspace{-6mm}
\end{figure}

Combat (Figure \ref{fig:combat}) enables direct agent-agent confrontation by implementing three different attack "styles":
 \begin{itemize}
     \setlength\itemsep{0pt}
     \item \textbf{Melee:} Inflicts 10 damage at 1 range
     \item \textbf{Ranged:} Inflicts 2 damage at 1-2 range
     \item \textbf{Mage:} Inflicts 1 damage at 1-3 range and freezes the target in place, preventing movement for two ticks
 \end{itemize}
 
Each point of damage inflicted steals one point of food and water from the target and returns it to the attacker. This serves as an incentive to engage in combat. It is still fully possible for agents to develop primarily foraging based strategies, but they must at least be able to defend themselves. The combat styles defined impose clear but difficult to optimize trade offs. Melee combat fells the target in one attack, but only if they are able to make their attack before the opponent retaliates in kind. Ranged combat produces less risky but more prolonged conflicts. Mage combat does little damage but immobilizes the target, which allows the attacker to retreat in favor of a foraging based strategy. More aggressive agents can use mage combat to immobilize their target before closing in for the kill. In all cases, the best strategy is not obvious, again imposing an arms race.

\textbf{Technical details:}
\begin{itemize}
    \setlength\itemsep{0pt}
    \item \textbf{Attack range} is defined by l1 distance: "1 range"  is a 3X3 grid centered on the attacker.
    \item \textbf{Spawn Killing} Agents are immune during their first 15 game ticks alive. This prevents an exploit known as "spawn killing" whereby players are repeatedly attacked immediately upon entering the game. Human games often contain similar mechanism to prevent this strategy, as it results in uninteresting play.
\end{itemize}

\section*{API}
The initial release is bundled with two APIs for running experiments on our platform. All of our experiments are RL based, but the API implementation is intentionally generic. Evolutionary methods and algorithmic baselines should work without modification.

\textbf{Gym Wrapper} We provide a minimal extension of the Gym VecEnv API \citep{DBLP:journals/corr/BrockmanCPSSTZ16} that adds support for variable numbers of agents per world and at any given time. This API distributes environment computation of observations and centralizes training and inference. While this standardization is convenient, MMOs differ significantly from arcade games, which are easier to standardize under a single wrapper. The Neural MMO setting requires support for a large, variable number of agents that run concurrently, with aggregation across many randomly generated environments. The Gym API incurs additional communications overhead that the native API bypasses.

\begin{figure}
\begin{center}
\resizebox{1.0\linewidth}{!}{
\includegraphics{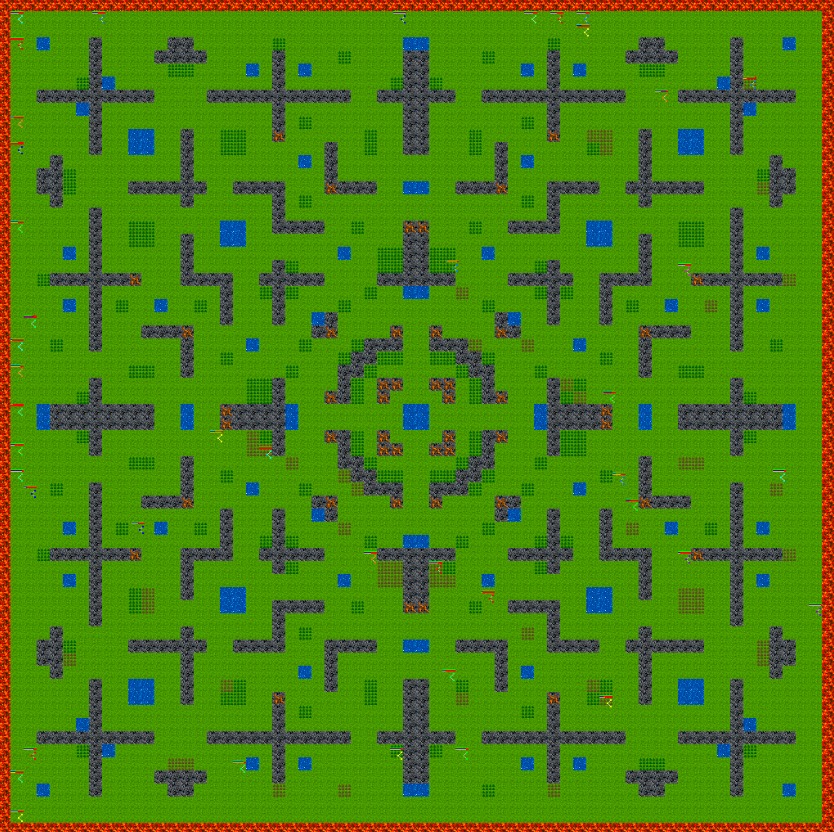}
}
\end{center}
\caption{Example map in the 2D client}
\label{fig:env2d}
\end{figure}

\begin{figure}
\begin{center}
\resizebox{1.0\linewidth}{!}{
\includegraphics{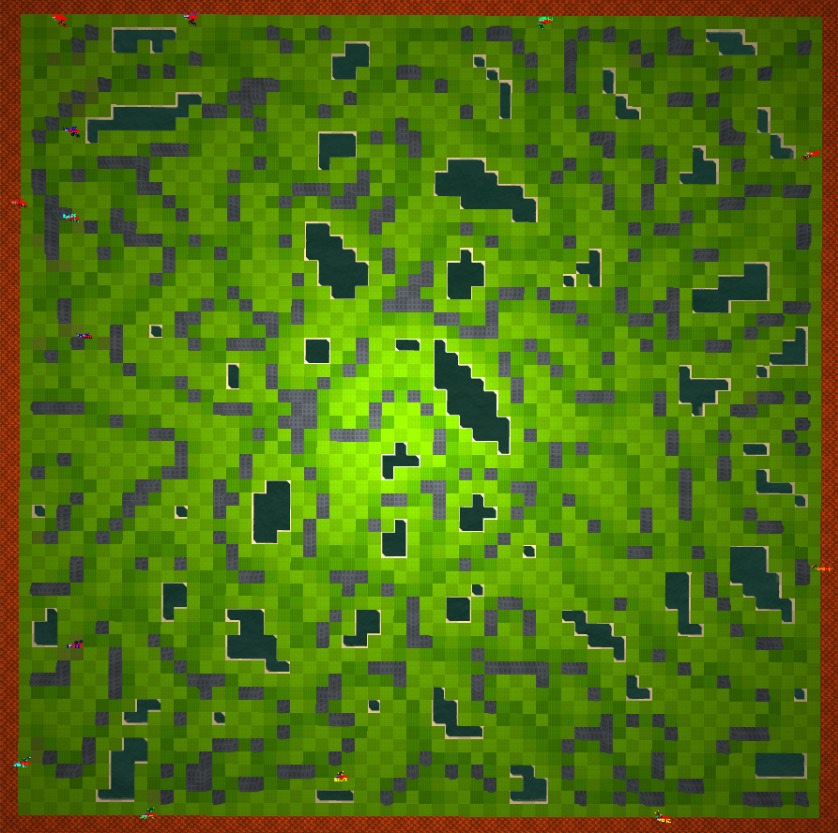}
}
\end{center}
\caption{Example overhead map view in the 3D client}
\label{fig:env3d}
\end{figure}

\textbf{Native} This is the simplest and most efficient interface. It pins the environment and agents on it to the same CPU core. Full trajectories run locally on the same core as the environment. Interprocess communication is only required infrequently to synchronize gradients across all environments on a master core. We currently do the backwards pass on the same CPU cores because our networks are small, but GPU is fully supported.

\begin{figure*}
\begin{center}
\resizebox{0.95\linewidth}{!}{
\includegraphics{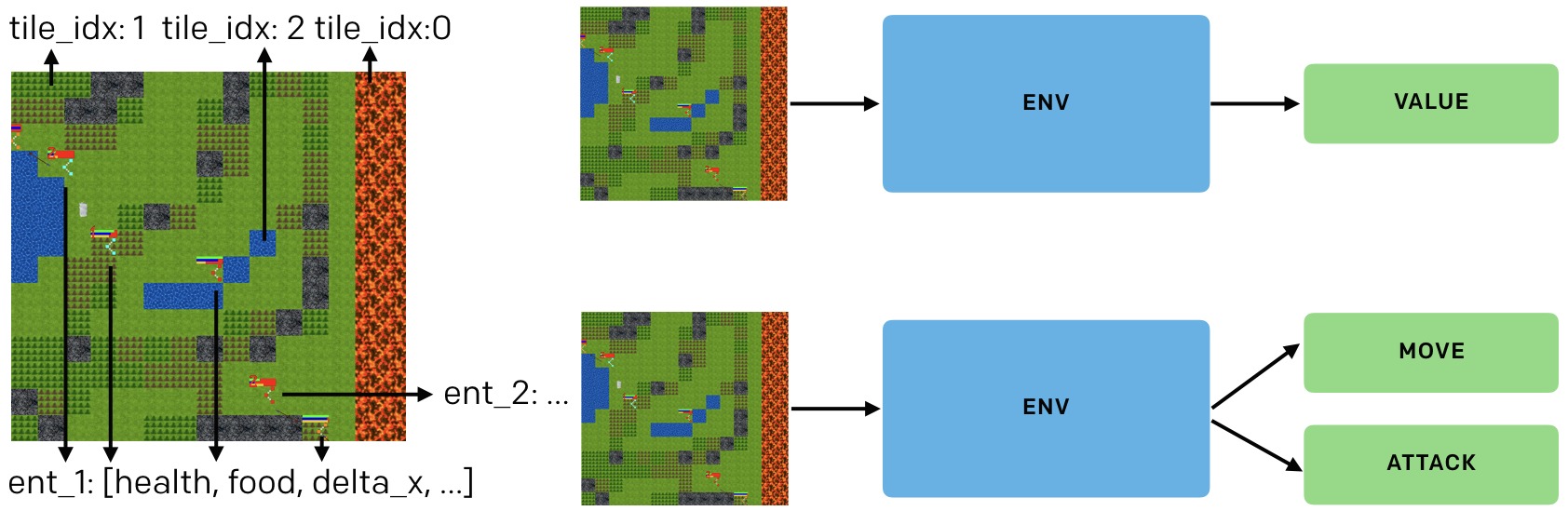}
}
\end{center}
\caption{Agents observe their local environment. The model embeds this observations and computes actions via corresponding value, movement, and attack heads. These are all small fully connected networks with 50-100k parameters.}
\label{architecture}
\end{figure*}

\section*{Client}
\begin{figure*}
\begin{center}
\resizebox{\linewidth}{!}{
\includegraphics{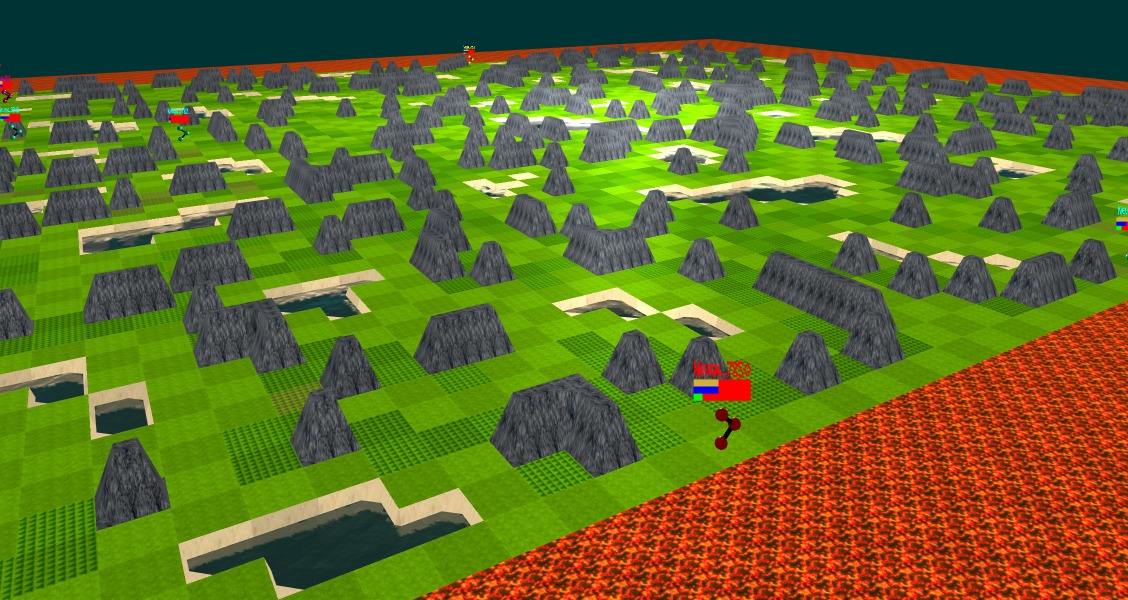}
}
\end{center}
\caption{Perspective screenshot of 3D environment}
\label{fig:envfullbleed}
\end{figure*}

The environment visualizer comes bundled with research tools for analyzing agent policies. In the initial release, we provide both a 2D python client (Figure \ref{fig:env2d}) and a 3D web client (Figure \ref{fig:env3d}, \ref{fig:envfullbleed}). The 3D client has the best support for visualizing agent policies. The 2D client is already deprecated; we include it only because it will likely take a few weeks to fully finish porting all of the research tools. We include visualization tools for producing the following visualization maps; additional documentation is available on the project Github:

\begin{multicols}{2}
\begin{itemize}
    \setlength\itemsep{0pt}
    \item Value ghosting
    \item Exploration
    \item Interagent dependence
    \item Combat
\end{itemize}
\end{multicols}

\section*{Policy training and architecture}
Parameters relevant to policy training are listed in Table 1. The neural net architecture, shown in Figure \ref{architecture}, is a simplest possible fully connected network. It consists of a preprocessor, main network, and output heads. The preprocessor is as follows: 

\begin{itemize}
    \setlength\itemsep{0pt}
    \item Embed indicies corresponding to each tile into a 7D vector. Also concatenates with the number of occupying entities.
    \item Flatten the tile embeddings
    \item Project visible attributes of nearby entities to 32D
    \item Max pool over entity embeddings to handle variable number of observations
    \item Concatenate the tile embeddings with the pooled entity embeddings
    \item Return the resultant embedding
\end{itemize}

The main network is a single linear layer. The output heads are also each linear layers; they map the output hidden vector from the main network to the movement and combat action spaces, respectively. Separate softmaxes are used to sample movement and combat actions.

Technical details
\begin{itemize}
    \setlength\itemsep{0pt}
    \item For foraging experiments, the attack network is still present for convenience, but the chosen actions are ignored.
    \item  Note that 1D max pooling is used to handle the variable number of visible entities. Attention \cite{DBLP:journals/corr/BahdanauCB14} may appear the more conventional approach, but recently \cite{OpenAI_dota} demonstrated that simpler and more efficient max pooling may suffice. We are unsure if this is true at our scale, but used max pooling nonetheless for simplicity.
\end{itemize}

\begin{table*}
\caption{Training details and parameters for all experiments}
\begin{tabular}{ c | c c }
 Parameter & Value & Notes \\ 
 \hline
 Training Algorithm & Policy Gradients\cite{williams1992simple} & + Value function baseline \\ 
 Adam Parameters & lr=1e-3 & Pytorch Defaults \\  
 Weight Decay & 1e-5 & Training stability is sensitive to this \\   
 Entropy Bonus & 1e-2 & To stabilize training; possibly redundant \\
 Discount Factor & 0.99 & No additional trajectory postprocessing \\
\end{tabular}
\end{table*}

\end{document}